\documentclass[12pt]{article}

\usepackage{epsfig}
\usepackage{Wrapfig}
\usepackage{cite}

{\end{list}}
\newcounter{enumct}

\newcommand{\lep}{{\mathrm l}} 
\newcommand{\e}{{\mathrm e}}

\newcommand{\q}{{\mathrm q}}

\newcommand{\qb}{\mathrm{\bar{q}}}

\newcommand{\p}{{\mathrm p}}
\newcommand{\X}{{\mathrm X}}
\newcommand{\K}{{\mathrm K}}
\newcommand{\epl}{{\mathrm e^{+}}}
\newcommand{\emi}{{\mathrm e^{-}}}
\newcommand{\mpl}{{\mathrm \mu^{+}}}
\newcommand{\mmi}{{\mathrm \mu^{-}}}
\newcommand{\lpl}{{\mathrm l^{+}}}
\newcommand{\lmi}{{\mathrm l^{-}}}
\newcommand{\pom}{\alpha_{I\!\!P}}

\def\lsim{\mathrel{\rlap{\lower4pt\hbox{\hskip1pt$\sim$}}
    \raise1pt\hbox{$<$}}}         
\def\gsim{\mathrel{\rlap{\lower4pt\hbox{\hskip1pt$\sim$}}
    \raise1pt\hbox{$>$}}}         

\begin{document}
 
\sloppy

\pagestyle{empty}

\renewcommand{\thefootnote}{\fnsymbol{footnote}}
\setcounter{footnote}{0}

\begin{center}
{\LARGE\bf The structure of the photon}\\[4mm]
{\LARGE\bf and its interactions\footnote{Invited talk given at the 
Workshop on Photon Interactions and the Photon Structure, Lund, Sweden, September 10-13, 1998.}}\\[10mm]
{\Large Bernd Surrow} \\[3mm]
{\it CERN, EP-division/OPAL}\\[1mm]
{\it CH-1211 Geneva 23, Switzerland}\\[1mm]
{\it E-mail: Bernd.Surrow@cern.ch}\\[20mm]

{\bf Abstract}\\[1mm]
\begin{minipage}[t]{140mm}
The OPAL experiment at LEP has performed a variety of measurements of 
$\gamma\gamma$ and $\e\gamma$ scattering at the $\epl\emi$ collider LEP
to gain a deeper insight into the structure of the photon and its
interactions. This review presents a summary of these results.
\end{minipage}\\[5mm]

\rule{160mm}{0.4mm}

\end{center}

\renewcommand{\thefootnote}{\arabic{footnote}}
\setcounter{footnote}{0}

\section{Introduction}

\begin{wrapfigure}{r}{8.2cm}
\vspace*{-0.8cm}
\setlength{\unitlength}{1.0cm}
\begin{picture} (15.0,2.5)
\put (0.25,1.25){{\small (a)}}
\put (3.0,1.25){{\small (b)}}
\put (5.75,1.25){{\small (c)}}
\put (0.25,0.0){{\small point-like}}
\put (5.0,0.0){{\small VM ($J^{PC}=1^{--}$)}}
\put (-0.25,0.5){\mbox{\epsfig{figure=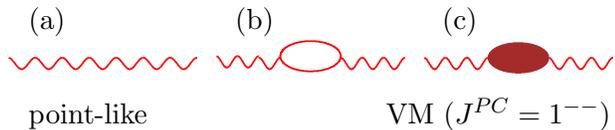,width=8.5cm,clip=}}}
\end{picture}
\caption{\it{Dual nature of the photon.}}
\label{fig1}
\end{wrapfigure}

\begin{wrapfigure}{r}{8.2cm}
\hspace*{1.0cm}
\epsfig{figure=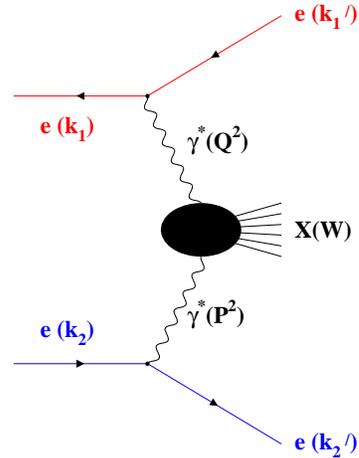,width=5.0cm,clip=}
\caption{\it{Kinematics of two-photon induced processes in $\epl\emi$ collisions:
$k_{1}$ ($k_{1}^{'}$) and $k_{2}$ ($k_{2}^{'}$) are the four-momenta of the incoming (outgoing) 
electrons. $q$ and $p$ are the four-momenta of the mediated virtual photons. 
$W$ is the invariant mass of the final state system~$\X$.}}
\label{fig2}
\end{wrapfigure}

The photon plays the fundamental role as the gauge 
boson of mediating electromagnetic interactions through the coupling to charged particles.
The photon appears to be in this respect as a point-like particle (Figure \ref{fig1} (a)). 
However, a photon is subject to quantum fluctuations, as denoted in Figure \ref{fig1}. 
It can fluctuate into a lepton/anti-lepton pair, $\lep\mathrm{\bar{l}}$, or a 
quark/anti-quark pair, $\q\qb$ (Figure \ref{fig1} (b) and (c)). 
If, e.g in $\gamma \p$ interactions, the fluctuation time is large compared to the
interaction time, $\gamma \p$ interactions can proceed through the interaction of 
a $\q\qb$ pair and the proton, which gives rise to a hadronic type reaction \cite{ref1_ioffe}.
This behavior has been incorporated in the vector dominance model (VDM) \cite{ref2_sakurai} which
describes $\gamma \p$ interactions as the interaction of the proton and a vector meson
resulting from a quantum fluctuation of the photon with the vector meson state
having the same quantum numbers as the photon, i.e. $J^{PC}=1^{--}$.
This dual nature of the photon, i.e. point-like and hadron-like, 
leads to a variety of phenomena which have been investigated
in $\gamma \p$ and $\gamma \gamma$/$\e\gamma$\footnote{Positrons are referred to as electrons
in this review unless otherwise stated.} scattering at several experiments at HERA and 
LEP, respectively \cite{ref3_erdmann}. The underlying
kinematics at the $\epl\emi$ collider LEP is shown in Figure \ref{fig2}. 
The initial state consists
of the incoming electron and positron, each of which - to lowest order 
in $\alpha_{\rm em}$ - emits a virtual photon. 
The square of the four momentum transfer or the virtuality of the mediated virtual photons 
is denoted by $Q^{2}=-q^{2}=(k_{1}-k_{1}^{'})^{2}$, which will be later referred to as the probe 
virtuality, and $P^{2}=-p^{2}=(k_{2}-k_{2}^{'})^{2}$ which will be later referred to as the 
target virtuality. $Q^{2}$ ($P^{2}$) 
can be reconstructed from the energy and angle of the respective detected (tagged) electron
as follows:
$Q^{2}=2E_{\rm beam}E_{\rm tag,1}(1-\cos\theta_{\rm tag,1})$ 
($P^{2}=2E_{\rm beam}E_{\rm tag,2}(1-\cos\theta_{\rm tag,2})$).
The invariant mass of the final state $\X$ is denoted
by $W=(q+p)^{2}$. The Bjorken scaling variable $x$ is defined as $x=Q^{2}/(Q^{2}+P^{2}+W^{2})$ and
the `inelasticity' as $y=(p \cdot q)/(p \cdot k)$. The inelasticity can be reconstructed
from the energy and angle of the tagged electron as follows: 
$y=1-E_{\rm tag,1}/E_{\rm beam}(\cos^{2}\theta_{\rm tag,1}/2)$.

Three event classes can be distinguished depending on whether the electron in the final
state is tagged or not. If both final state electrons are not tagged, both
virtualities $Q^{2}$ and $P^{2}$ - depending on the detector acceptance - are small,
and the photons can be considered as quasi-real, i.e. $Q^{2} \simeq 0 $ and $P^{2} \simeq 0$.
This is the case of anti-tagged events which allows
to investigate $\gamma\gamma$ scattering and thus photon interactions. 
If one of the scattered electrons,
is detected (tagged), the process shown in Figure~\ref{fig2} 
can be considered as the scattering of an electron 
off a quasi-real photon. In this case of $e\gamma$ scattering (single-tagged events) 
one is able to study 
the photon structure similarly to the case of lepton-nucleon scattering 
($Q^{2} \gg P^{2}$). The case in which both
final state electrons are tagged is referred to as double-tagged events. 

Unlike to $\e\p$ and $\gamma \p$ scattering e.g. at HERA, the invariant mass $W$ 
and therefore $x$ in $\e\gamma$ and $\gamma \gamma$ scattering at LEP
can only be reconstructed from the final state $\X$.
In case of $\X$ being a hadronic final state, contrary to the case of lepton pairs such as
$\mpl\mmi$ pairs,
the reconstruction of $W$ suffers from particle losses due to a limited detector acceptance. 
This gives
rise to the fact that the measured invariant mass $W_{\rm vis}$ is different from the 
invariant mass $W$. 
One therefore needs to unfold the true result from the visible measurement. This requires a good 
Monte Carlo (MC) simulation of the underlying event structure. This will be 
discussed in more detail in section 2.2.
Only double-tagged events permit a reconstruction of $W$ independent of the hadronic final state
through the reconstruction of both photon virtualities $Q^{2}$ and $P^{2}$ from the tagged
final state electrons.

The OPAL experiment at LEP has performed a variety of measurements of single-tagged 
and anti-tagged events to gain a deeper insight into the structure of the photon and its
interactions. The following review will provide a summary of various measurements of those two 
event classes. This includes in the case of single-tagged events the investigation of the 
photon structure through the extraction of the 
photon structure function for the case of hadronic final states and a final state which 
consists of a $\mpl\mmi$ pair as well as the study of azimuthal correlations of 
$\mpl\mmi$ pairs. Anti-tagged
events have been used to study the production of jets and hadrons in $\gamma\gamma$ scattering as
well as for a measurement of the total hadronic $\gamma\gamma$ cross-section. 
Results from these measurements will be presented in detail in section 2 for single-tagged events 
and section 3 for anti-tagged events. Section 4 provides a short summary and an outlook of 
further results on the structure of the photon and its interactions to be expected from the OPAL 
experiment at LEP.

\section{Electron-Photon scattering}

\subsection{General considerations}

The starting point to study $\e\gamma$ scattering and thus the structure of the photon
in a similar way as $\e\p$ scattering at LEP, is the process 
$\e \e \rightarrow \e \e \X$.
It can be viewed as deep-inelastic scattering of an electron on a quasi-real photon. 
This requires to transform the cross-section $d\sigma_{\e \e \rightarrow \e \e \X}$ 
into an $\e\gamma$ cross-section, 
$d\sigma_{\e \gamma \rightarrow \e \X}$:
\begin{equation}
d\sigma_{\e\e \rightarrow \e\e \X} = d\sigma_{\e\gamma\rightarrow \e \X} \cdot f_{\gamma/\e}
\end{equation}
The flux factor $f_{\gamma/\e}$, which denotes the flux of the target photon off the 
incoming electron, takes
into account the momentum spread of the target photon. This is in contrast to $\e \p$ 
scattering where the energy of both incoming particles is known. 
The calculation of the flux factor $f_{\gamma/\e}$ is carried out using the 
equivalent photon approximation (EPA) \cite{ref4_weiz,ref5_budnev}. 
Provided that the target photon is almost real, i.e. 
$P^{2}\simeq 0$, 
and thus almost solely transversally polarized, the $\e \gamma$ cross-section can be 
reduced to two components only, 
$\sigma_{TT}$ and $\sigma_{LT}$. The first index denotes the polarization of the 
probing
photon whereas the second index refers to the polarization of the target photon. 
The following relations are defined:
\begin{eqnarray}
F_{1}^{\gamma}(x,Q^{2}) & = & \frac{Q^{2}}{4\pi^{2}\alpha}\frac{1}{2x} \cdot \sigma_{TT} \nonumber \\
F_{2}^{\gamma}(x,Q^{2}) & = & \frac{Q^{2}}{4\pi^{2}\alpha} \cdot (\sigma_{TT}+\sigma_{LT})
\end{eqnarray}
$F_{1,2}^{\gamma}$ are denoted as the photon structure functions.
$F_{L}^{\gamma}=F_{2}^{\gamma}-2xF_{1}^{\gamma}$ is the longitudinal structure function.
Using these relations, the 
following expression for the cross-section of deep-inelastic $\e\gamma$ scattering is obtained:
\begin{equation}
\frac{d^{2}\sigma(\e\gamma\rightarrow \e \X)}{dxdQ^{2}}=\frac{2\pi\alpha^{2}}{xQ^{4}}\left[(1+(1-y)^{2})F_{2}^{\gamma}(x,Q^{2})-y^{2}F_{L}^{\gamma}(x,Q^{2})\right]
\end{equation}
with $\alpha$ being the fine structure constant. 
For energies of the tagged electron larger than half the beam energy, which is used to reject
beam-associated background events, $y$ is much less than $1$ and thus the contribution of 
$y^{2}F_{L}^{\gamma}$
to $d^{2}\sigma(\e\gamma\rightarrow e \X)/dxdQ^{2}$ is small and can therefore be neglected. 
In this case, $d^{2}\sigma(\e\gamma\rightarrow e \X)/dxdQ^{2}$ is directly proportional to 
$F_{2}^{\gamma}$. 

The above expression for $d^{2}\sigma(\e\gamma\rightarrow \e \X)/dxdQ^{2}$  
is deduced for the case of transversally polarized target photons, i.e.
only the two components $\sigma_{TT}$ and $\sigma_{LT}$ are taken into account. In case of virtual
target photons ($P^{2}\neq 0$), 
i.e. in the case of double-tagged events, this simplification no longer holds and additional 
cross-section terms have to be taken into account. This will change in particular the contribution
of $F_{2}^{\gamma}$ to $d^{2}\sigma(\e\gamma\rightarrow \e \X)/dxdQ^{2}$ 
\cite{ref5_budnev,ref9_pluto}. 

For a final state $\X$ which consists of lepton pairs (described by QED) the 
photon structure functions 
are referred to as $F_{1,2, {\rm QED}}^{\gamma}$ (leptonic structure functions), 
whereas for a hadronic final state $\X$ created by a $\q\qb$ pair (described by QCD), 
the photon structure functions are denoted as $F_{1,2}^{\gamma}$ (hadronic structure functions). 
$F_{2, {\rm QED}}^{\gamma}$ is predicted by QED and a measurement of 
$F_{2, {\rm QED}}^{\gamma}$ serves as a test of QED. $F_{2}^{\gamma}$ however 
cannot be calculated in a similar way from first principles for all $x$ and $Q^{2}$. 
Several models have been developed 
based on perturbative QCD (pQCD) and a particular non-perturbative QCD ansatz.
A measurement of $F_{2}^{\gamma}$ thus allows to investigate QCD.

Despite the similarity of the differential cross-sections for $\e \gamma$ scattering
and $\e \p$ scattering, there are subtle differences in the behavior of the structure functions
which will be briefly summarized in the following.

As pointed out in the previous section, the photon shows in its interaction contributions
from a point-like as well as a hadron-like behavior. The lowest-order point-like
contribution, i.e. the purely electromagnetic process $\gamma^{*}\gamma \rightarrow \q \qb$, 
can already be predicted in the Quark-Parton model (QPM) and allows therefore a prediction of the photon
structure function. 
The hadronic contribution cannot
be calculated from first principles similarly to the case of the proton. One relies on a 
non-perturbative QCD input such as an estimate within the framework of the
vector dominance model (VDM). 
$F_{2}^{\gamma}$ is large for high values of $x$, whereas the proton structure function $F_{2}^{p}$
decreases at large $x$. Furthermore, $F_{2}^{\gamma}$ increases with $Q^{2}$ at all values
of $x$, which is expected already from the QPM, i.e. from the point-like
contribution and thus without the presence of gluon radiation. This is in striking contrast
to the $Q^{2}$ dependence of $F_{2}^{p}$. 

\begin{wrapfigure}{r}{8.2cm}
\hspace*{0.5cm}
\epsfig{figure=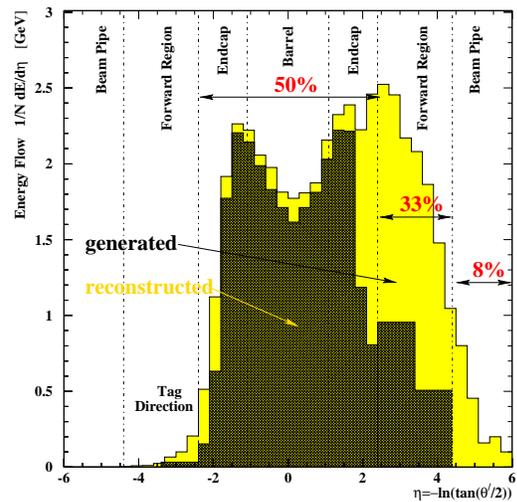,width=7.0cm,clip=}
\caption{\it Hadronic energy flow per event $(1/N) \cdot dE/d\eta$ as a 
function of pseudorapidity $\eta$ obtained from the HERWIG MC generator.}
\label{fig4}
\end{wrapfigure}

The description of the photon structure functions in terms of parton distributions 
and their $Q^{2}$ evolution based on the DGLAP equations can be developed in a similar
way as for the proton. The point-like contribution due to the splitting
process $\gamma^{*}\gamma \rightarrow \q\qb$ 
can be incorporated as an additional
splitting function besides the usual QCD splitting functions. This additional term
results in the fact that the DGLAP equations are no longer homogeneous. The solution
of the inhomogeneous part is determined by the photon splitting function due
to the splitting process $\gamma^{*}\gamma \rightarrow \q\qb$ whereas 
the homogeneous solution obeys the hadron-like evolution of the DGLAP equations.

In summary, the hadronic photon structure function $F_{2}^{\gamma}$ is given as the sum
of a point-like, $F_{2}^{\gamma,{\rm pl}}$, and a hadron-like, $F_{2}^{\gamma,{\rm hadr.}}$, 
contribution:
\begin{equation}
F_{2}^{\gamma}=F_{2}^{\gamma,{\rm pl}}+F_{2}^{\gamma,{\rm hadr.}}
\end{equation}
At large $x$ and asymptotically large $Q^{2}$, the value of $F_{2}^{\gamma, {\rm hadr.}}$ can be 
calculated
from pQCD. The NLO result obtained by Bardeen and Buras \cite{ref11_bardeen} is given as 
follows:
\begin{equation}
\frac{F^{\gamma}_{2}}{\alpha}=\frac{a(x)}{\alpha_{s}(Q^{2})}+b(x)
\end{equation}
$a(x)$ and $b(x)$ are calculable functions which diverge for $x\rightarrow 1$. $\alpha_{s}(Q^{2})$
is the strong coupling constant. The first term reflects the LO result by Witten 
\cite{ref12_witten}. The $Q^{2}$ evolution of $F_{2}^{\gamma, {\rm hadr.}}$ will eventually
allow to extract the QCD scale $\Lambda_{{\rm QCD}}$ provided that the 
presence of non-perturbative contributions are under control.

Various models have been developed in the past on the description of the photon structure function
such as the model by Gl\"{u}ck, Reya and Vogt (GRV) \cite{ref_grv} and the model by Schuler and 
Sj\"{o}strand (SaS) \cite{ref_sas}.
Common to these descriptions is a non-perturbative ansatz for the hadronic contribution at a 
starting scale $Q^{2}_{0}$ and a subsequent DGLAP evolution, including a 
particular treatment of the charm contribution. They differ in the way this procedure is
carried out. Section 2.3 will discuss in more detail these models by comparing them to 
experimental results on $F_{2}^{\gamma}$. 

\subsection{Hadronic energy flows}

The measurement of the hadronic structure function $F_{2}^{\gamma}$ 
requires the determination of the invariant mass $W$ from the hadronic final state as 
pointed out in section~1.
The need to unfold the invariant mass, $W$, from the measured result, $W_{\rm vis}$,
requires a modeling of the hadronic final state incorporated in a particular Monte Carlo (MC) 
generator. The 
measurement of $F_{2}^{\gamma}$ and the simulation of the hadronic final state are
thus closely connected. 

\begin{wrapfigure}{r}{8.2cm}
\hspace*{0.5cm}
\vspace*{-0.5cm}
\epsfig{figure=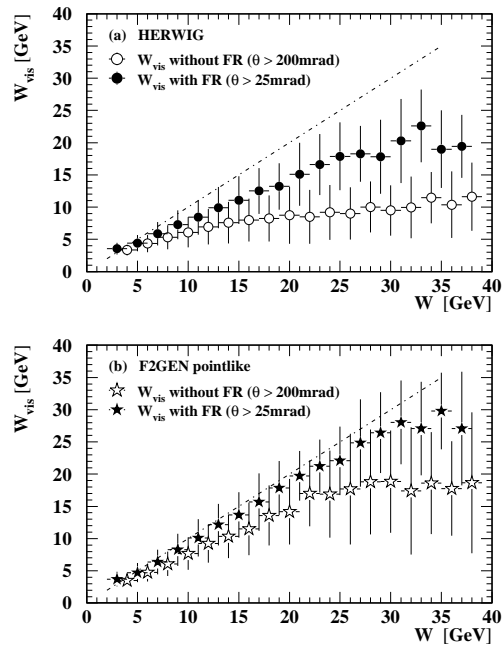,width=7.0cm,clip=}
\caption{\it Correlation between the observed
hadronic invariant mass $W_{\rm vis}$ and the generated hadronic invariant mass $W$.} 
\label{fig5}
\end{wrapfigure}

Several investigations have been carried out by OPAL \cite{ref14_opal}, 
comparing various observables such as the hadronic
energy flow between data and MC to gain a deeper insight into the simulation of the hadronic final state.
The two general purpose QCD based MC generators HERWIG \cite{ref15_herwig} and PYTHIA \cite{ref16_pythia} 
have been used for these studies. In addition, the F2GEN generator \cite{ref17_twogen} has been 
applied to study
various aspects of the simulation of a $\q\qb$ final state in the $\gamma^{*}\gamma$ centre-of-mass
system. The F2GEN `point-like' mode represents the unphysically extreme case of a two-quark state
in the $\gamma^{*}\gamma$ centre-of-mass system with an angular distribution as in lepton pair production
from two real photons. The F2GEN `permiss' mode reflects a physics motivated mixture of point-like 
and peripheral interactions, where peripheral means that the transverse momentum of outgoing quarks 
is given by an exponential distribution as if the photons interacted as pre-existing hadrons and direct
photon-quark couplings never occurred. 

The hadronic system $\X$ is strongly boosted along the beam direction, accentuating the loss of
particles within the well-measured region. Figure \ref{fig4} shows the prediction of the HERWIG
MC generator for the hadronic energy flow per event, $(1/N) \cdot dE/d\eta$, as a function of 
the pseudorapidity $\eta=-\ln\,\tan(\theta/2)$. The tagged electron is not included in the energy flow.
By definition, it lies in the 
negative rapidity region $-3.5 < \eta < -2.8$. The generated energy flow is shown as the light 
shaded region. The dark shaded region shows the energy flow after reconstruction by the OPAL
detector including a simulation of the detector response. A significant fraction of the energy flow
goes into the forward direction. Approximately $50\%$ is deposited in the central region whereas
$33\%$ are deposited in the forward region. Only $8\%$ of the total hadronic energy flow lies outside 
the detector acceptance. 
Figure \ref{fig5} shows the
correlation between the generated hadronic invariant mass $W$ and the observed
hadronic invariant mass $W_{\rm vis}$ with and without sampling the hadronic energy using the
OPAL forward detectors (FD). Including a measurement of the hadronic energy in the forward
direction by sampling the hadronic energy using the OPAL forward detectors substantially
improves the correlation between $W$ and $W_{\rm vis}$. The degree of correlation 
depends on the MC model.

\begin{figure}[t]
\setlength{\unitlength}{1.0cm}
\begin{picture} (15.0,6.0
)
\put (1.25,0.0){\mbox{\psfig{figure=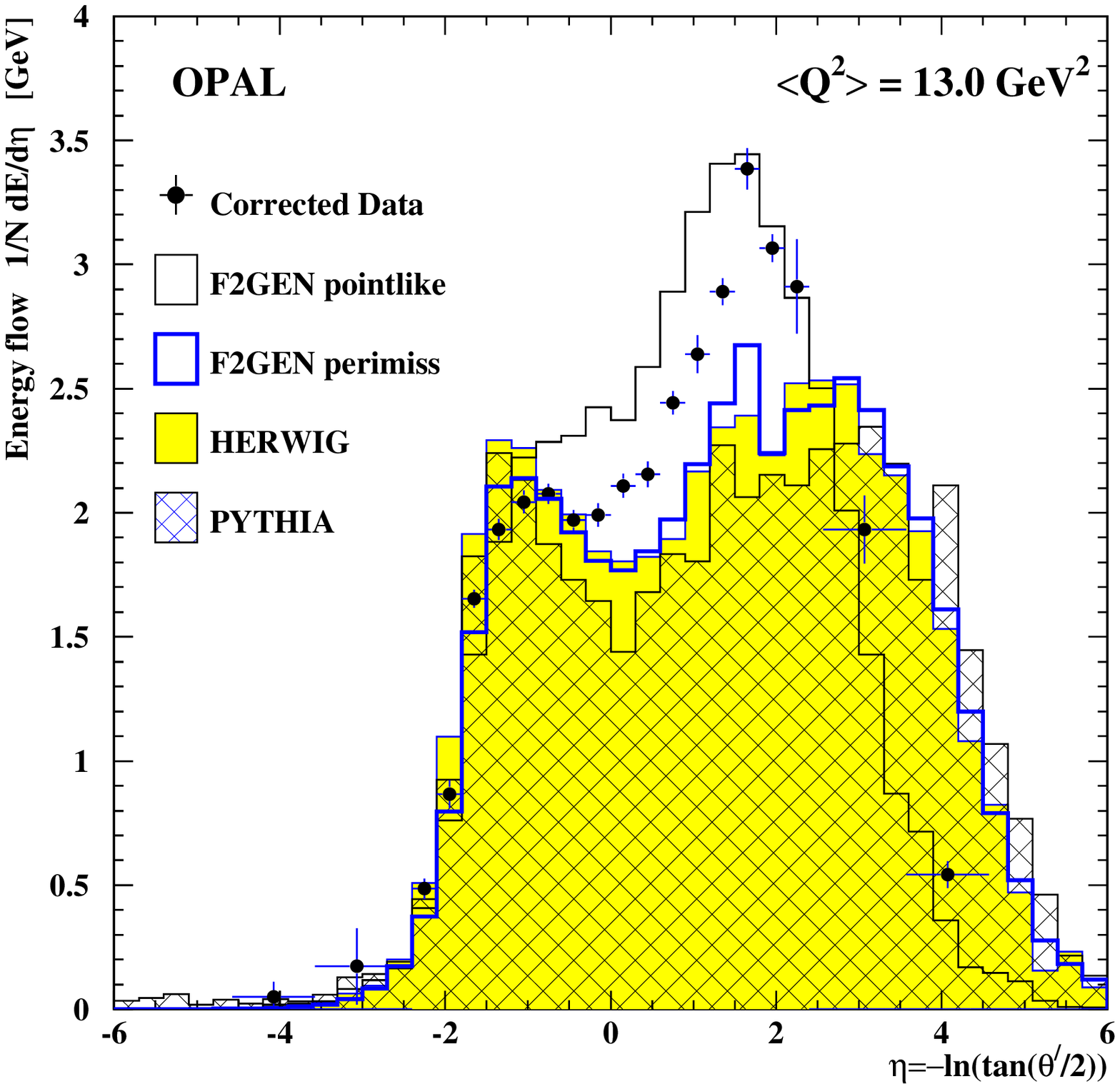,width=6.0cm,clip=}}}
\put (8.75,0.0){\mbox{\psfig{figure=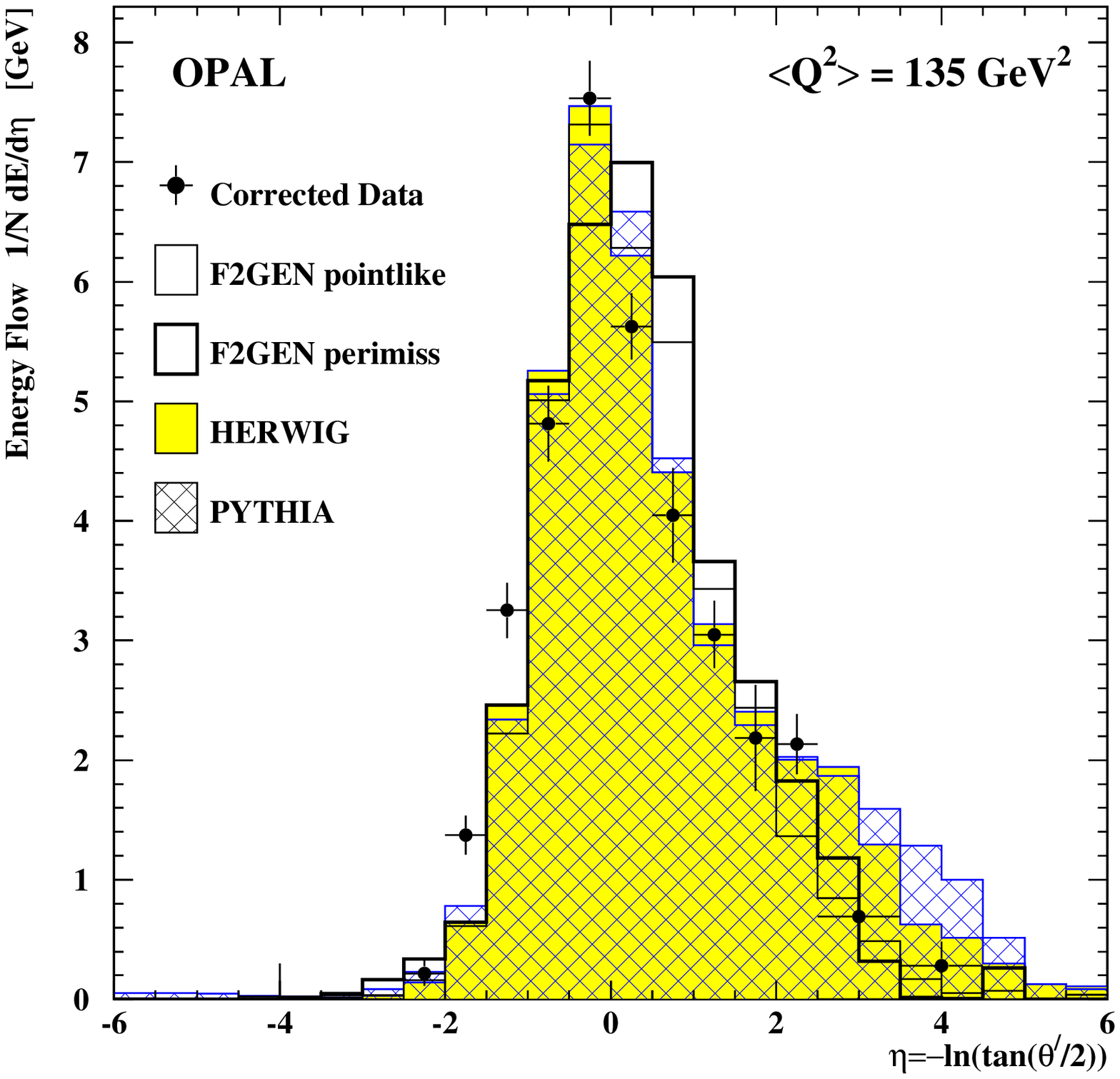,width=6.0cm,clip=}}}
\end{picture}
\caption{\it Hadronic energy flow per event as a function of pseudorapidity $\eta$ for 
low $Q^{2}$ ($Q^{2}=13.0\,$GeV$^{2}$) and high $Q^{2}$ ($Q^{2}=135\,$GeV$^{2}$) events. 
The data distributions have been corrected for detector effects.}
\label{fig7}
\end{figure}

The comparison between data and MC distributions of the hadronic energy flow is shown in Figure 
\ref{fig7} as a function of the pseudorapidity $\eta$ for two values of $Q^{2}$ of 
$13.0\,$GeV$^{2}$ and $135\,$GeV$^{2}$. The data distribution has been corrected for detector 
effects. Large differences are seen between data and all MC models both within the central
region ($|\eta|<2.3$), where the energy is well measured and the forward region where the
energy is only sampled. The discrepancy increases towards small value of $Q^{2}$ and thus towards
small values of $x$. 

\begin{wrapfigure}{r}{8.2cm}
\epsfig{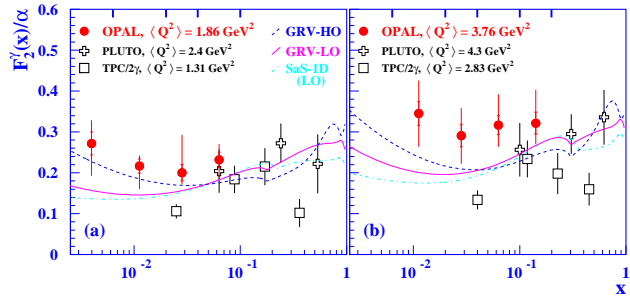}
\caption{\it OPAL results on 
$F_{2}^{\gamma}/\alpha$ as a function of $x$ for $Q^{2}=1.86\,{\rm GeV}^{2}$ 
and $Q^{2}=3.76\,{\rm GeV}^{2}$ together with results from other experiments.}
\label{fig8}
\end{wrapfigure}

It is apparent from this section that unfolding the distribution 
$W$ from the observed 
distribution $W_{{\rm vis}}$ will result in large systematic errors on the extracted photon 
structure function
$F_{2}^{\gamma}$ as long as the hadronic energy flow between the different MC 
models and 
the hadronic energy flow in data remains in clear disagreement. This is expected to affect in 
particular
the low $Q^{2}$ and low $x$ region. 

Better general purpose MC generators are required to provide a better description of the hadronic 
energy flow in data
and thus to reduce the large systematic uncertainties of the extracted photon structure function 
$F_{2}^{\gamma}$ (Section 2.3) \cite{ref18_leif}. 
These questions are being addressed
in a LEP-wide working group with the aim of a common and consistent presentation of the hadronic energy flow 
in data. An improvement 
of general purpose MC generators based on such investigations is eagerly awaited. It  
is only then that an improvement in the systematic uncertainties due to the description of the hadronic 
final state in the extraction of $F_{2}^{\gamma}$ can be expected.  

\begin{wrapfigure}{r}{8.2cm}
\epsfig{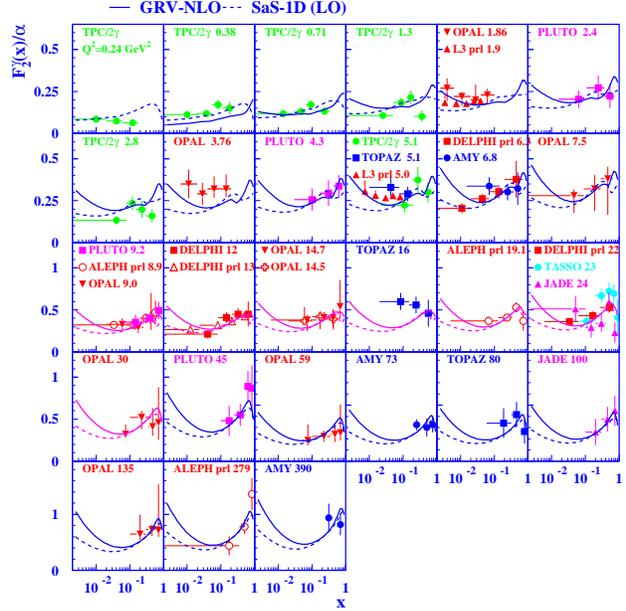}
\caption{\it OPAL results on 
$F_{2}^{\gamma}/\alpha$ as a function of $x$ for several $Q^{2}$ values 
together with results from other experiments.}
\label{fig9}
\end{wrapfigure}

\subsection{Hadronic structure function \boldmath{$F_{2}^{\gamma}$}}

The measurement of the hadronic structure function $F_{2}^{\gamma}$ has been
carried at OPAL as a function of $x$ and $Q^{2}$ over a wide kinematic range of
$0.0025<x \lsim~1$
and $1.1\,{\rm GeV}^{2}<Q^{2}<400\,{\rm GeV}^{2}$. Details on the 
analysis to extract $F_{2}^{\gamma}$ can be found in \cite{ref14_opal,ref19_opal}. 
The $Q^{2}$ dependence is expected to be logarithmic within the framework of pQCD.
The $x$ dependence of the proton structure function $F_{2}^{p}$ 
has been studied in detail at HERA
which shows a steep rise towards small values of $x$ at not too low $Q^{2}$ 
values \cite{ref37_zeus,ref20_hera}.
If the photon were purely hadron-like, a similar rise of $F_{2}^{\gamma}$ is 
expected.
A measurement of $F_{2}^{\gamma}$ as a function of $x$ and $Q^{2}$ therefore 
allows to shed light on the structure and the underlying dynamics of the photon as pointed 
out in section 2.1.

Figure \ref{fig8} shows OPAL results on $F_{2}^{\gamma}$ as a function
$x$ at low values of $Q^{2}$. OPAL results on 
$F_{2}^{\gamma}$ together with other experimental results are shown in Figure \ref{fig9}.
The full error bars show the statistical and systematic errors added in quadrature. 
In Figure \ref{fig8}, the statistical errors are displayed by horizontal lines across the 
respective errors bars. The precision of the measurement is dominated by systematic 
uncertainties due to 
the modeling of the hadronic final state as pointed out in detail in the previous section. 
As can be seen from Figure \ref{fig8}, the OPAL results at the two lowest $Q^{2}$ values
agree within errors with 
the published results from PLUTO. However, the results both from PLUTO and OPAL are higher and 
different in shape than the previous measurement from TPC in a similar kinematic
region.
Taking into account the large uncertainties, $F_{2}^{\gamma}$ is found 
to be rather flat (Figure \ref{fig8}) although a small rise towards low $x$
cannot be excluded. 
It can be seen from Figures \ref{fig8} and \ref{fig9} that 
$F_{2}^{\gamma}$ rises smoothly towards large $x$. This behavior is reasonably
well described by the GRV and SaS models. It can be seen from Figure \ref{fig8}
that the higher-order GRV prediction (GRV-HO) follows the data more closely compared
to the leading-order GRV prediction (GRV-LO). 

\begin{wrapfigure}{r}{8.2cm}
\hspace*{0.25cm}
\epsfig{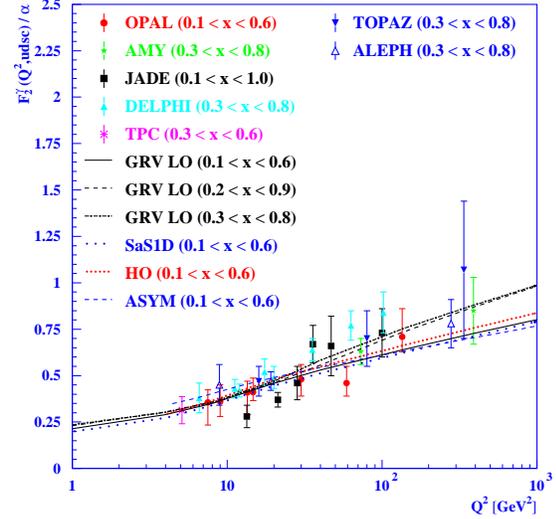}
\caption{\it OPAL results on 
$F_{2}^{\gamma}/\alpha$ as a function of $Q^{2}$ for different ranges in $x$
together with results from  other experiments \cite{nisius}.}
\label{fig10}
\end{wrapfigure}

The $Q^{2}$ dependence of $F_{2}^{\gamma}$ is shown in Figure \ref{fig10}
for the currently available data for four active flavors. It should be noted that
the quoted $x$ ranges by different experiments are not the same which makes a comparison
rather difficult since several predictions for different ranges in $x$
show a large difference for $Q^{2}>100\,$GeV$^{2}$.
All results on $F_{2}^{\gamma}$ as shown in Figure \ref{fig10} 
agree reasonably well, taking into account the large uncertainties. A logarithmic rise
of $F_{2}^{\gamma}$ is seen in the data as predicted by pQCD. 
However, the systematic uncertainties on $F_{2}^{\gamma}$ are too large
to perform a precision test of pQCD.
Both, the GRV and SaS model allow to describe the data equally well. The observed
logarithmic rise of $F_{2}^{\gamma}$ in $Q^{2}$ can be reasonably 
well represented by the pQCD leading order asymptotic solution. 
It is found that the hadronic contribution to  
$F_{2}^{\gamma}$ decreases towards larger $x$ and $Q^{2}$ values, and
amounts to only about $15\%$ at $Q^{2}=59\,$GeV$^{2}$ and $x=0.5$. 

All OPAL results on $F_{2}^{\gamma}$ have not been corrected for the fact
that in the single-tagged mode, $P^{2}$ is only
approximately zero. Several theoretical predictions exit on how $F_{2}^{\gamma}$
varies as a function of $P^{2}$ \cite{ref_p2}. 
It has been estimated based on the SaS model 
that $F_{2}^{\gamma}$ with $P^{2}\neq 0$ would be approximately
$10\%$ lower compared to the case of $P^{2}=0$. The $P^{2}$ distribution in the data
and the correct theoretical prescription are not known. This is reason why
no correction has been applied to the data on $F_{2}^{\gamma}$.
The larger centre-of-mass energy at 
LEP compared to former experiments at PETRA makes this effect to become more important.
The investigation of the virtual structure of the photon through double-tag events
and the measurement of $F_{2, {\rm QED}}^{\gamma}$ (Section 2.4)
will allow to gain some experimental knowledge in that respect. 

Improvements in the precision of $F_{2}^{\gamma}$ require considerable
improvements of general purpose MC models to 
better describe the hadronic final state. It is only
then that the study of the low $x$ behavior of $F_{2}^{\gamma}$ as well as 
as the $Q^{2}$ evolution could be carried out with higher precision. 

\begin{wrapfigure}{r}{8.2cm}
\vspace*{-0.25cm}
\hspace*{0.25cm}
\epsfig{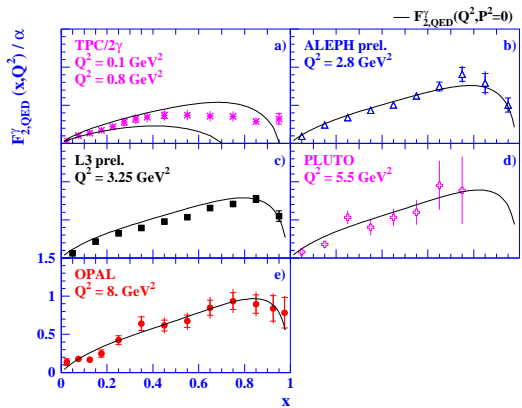}
\caption{\it $F_{2, {\rm QED}}^{\gamma}/\alpha$ as a function of $x$ \cite{nisius}.}
\label{fig11}
\end{wrapfigure}

\subsection{Leptonic structure function \boldmath{$F_{2, {\rm QED}}^{\gamma}$}}

The differential cross-section (3) with 
$\gamma^{*}\gamma \rightarrow \lpl \lmi$
allows to measure the leptonic structure function of the photon, 
$F_{2, {\rm QED}}^{\gamma}$.
A measurement of $F_{2, {\rm QED}}^{\gamma}$ has been carried out at OPAL 
by reconstructing $\mpl\mmi$ pairs
in the final state \cite{ref23_opal}. 
Such a clean environment allows to unfold the invariant mass
$W$ from the observed invariant mass $W_{\rm vis}$ with higher precision compared
to the case of a hadronic final state which yields a much more precise determination of 
the leptonic photon structure function compared to the hadronic case.

Figure \ref{fig11} shows OPAL results on
$F_{2, {\rm QED}}^{\gamma}$ as a function of $x$ together with other experiments.
The full error bars show the statistical and systematic errors added in quadrature, whereas
the statistical errors are displayed by horizontal lines across the 
respective errors bars. The precision of the data allow for a test of QED. The solid
lines in Figure~\ref{fig11} show the results of a QED calculation. All measurements
are consistent with QED expectations. The data are so precise that subtle effects 
such as the $P^{2}$ dependence can be investigated in more 
detail.

\begin{wrapfigure}{r}{8.2cm}
\hspace*{0.5cm}
\epsfig{figure=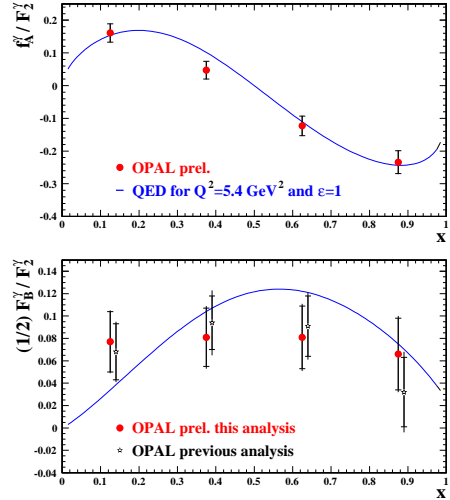,width=6.0cm,clip=}
\caption{\it OPAL results on the ratios $f_{A}^{\gamma}/F_{2}^{\gamma}$ and
$(1/2)F_{B}^{\gamma}/F_{2}^{\gamma}$
as a function of $x$.}
\label{fig13}
\end{wrapfigure}

\subsection{Azimuthal correlations of lepton pairs}

It has been pointed out that azimuthal correlations in the final-state particles of 
two photon collisions, i.e. $\gamma^{*}\gamma \rightarrow \X$, are sensitive to additional 
structure functions \cite{ref13_lep2}.
Azimuthal correlations can therefore supplement the direct measurement of the photon 
structure functions. 

OPAL has performed a measurement of azimuthal correlations in single-tagged 
events of $\gamma^{*}\gamma \rightarrow \mpl \mmi$ \cite{ref26_opala}. Two angles are defined 
which allow azimuthal correlations and thus more subtle structure functions to be 
studied. 
The azimuthal angle $\chi$ is the angle between the planes defined by
the $\gamma^{*}\gamma$ axis and the two-body final state and the 
$\gamma^{*}\gamma$ axis and the tagged electron. The second angle 
$\eta=\cos\theta^{*}$ is defined by the angle $\theta^{*}$ as the angle between the 
$\mmi$ and the $\gamma^{*}\gamma$ axis.

With the assumption that the target photon is only transversely polarized, the 
cross-section for $\e\gamma \rightarrow \e \mpl \mmi$ differential in $x$, $y$
and the two angles $\chi$ and $\eta$ is given as follows:
\begin{equation}
\frac{d\sigma(\e\gamma\rightarrow e\mpl\mmi)}{dxdyd\eta d\chi/2\pi} \approx \frac{2\pi\alpha^{2}}{Q^{2}}\left(\frac{1+(1-y)^{2}}{xy}\right) \left[2x\tilde{F}^{\gamma}_{T}+\tilde{F}_{L}^{\gamma}-\tilde{F}_{A}^{\gamma}\cos\chi+\frac{\tilde{F}_{B}^{\gamma}}{2}\cos 2\chi\right]
\end{equation}
The conventional structure functions can be recovered by integrating over 
$\chi$ and $\eta$:
\begin{equation}
F_{T,L,A,B}^{\gamma} = \frac{1}{2\pi}\int_{-1}^{+1}d\eta\int_{0}^{2\pi}d\chi \tilde{F}_{T,L,A,B}^{\gamma}
\end{equation}
In leading order and for massless muons, 
the following identity holds, $F_{B}^{\gamma}=F_{L}^{\gamma}$ although $F_{B}^{\gamma}$
and $F_{L}^{\gamma}$ are due to different helicity states of the photon. The
structure function $F_{A}^{\gamma}$ results from interference terms between
longitudinal and transverse photons whereas the structure function $F_{B}^{\gamma}$
refers to interference terms between solely transverse polarized photons. 
Figure \ref{fig13} shows published and preliminary results on the $x$ dependence of the ratios 
$f_{A}^{\gamma}/F_{2}^{\gamma}$ and $(1/2)F_{B}^{\gamma}/F_{2}^{\gamma}$ where
$f_{A}^{\gamma}=(1/2)(f_{A}^{+}-f_{A}^{-})$ with 
$f_{A}^{+}=(1/2\pi)\int_{0}^{+1}d\eta\int_{0}^{2\pi}d\chi \tilde{F}_{A}^{\gamma}$
and $f_{A}^{-}=(1/2\pi)\int_{-1}^{0}d\eta\int_{0}^{2\pi}d\chi \tilde{F}_{A}^{\gamma}$. 
The solid
line is the result of a QED prediction. The observed variation in $x$ of the
data is consistent with QED. Both ratios are significantly different from zero. 
These results not only serve as an interesting approach to supplement 
the conventional structure function measurements in itself, they also mark the first step
to perform such a measurement in the much more complex environment
of a hadronic final state.

\section{Photon-Photon scattering}

\subsection{Introduction}

Anti-tagged events in $\epl\emi$ scattering at LEP allow to study collisions of two quasi-real photons,
i.e. $\gamma\gamma \rightarrow \X$ with $Q^{2} \simeq 0 $ and $P^{2} \simeq 0$ as introduced
in section 1. The median $Q^{2}$ ($P^{2}$) amounts to approximately $10^{-4}\,$GeV$^{2}$ for 
$Q^{2}<4\,$GeV$^{2}$.
The $\epl\emi$ collider LEP is in this respect a $\gamma\gamma$ collider with a centre-of-mass
energy of the $\gamma\gamma$ system of $10<W_{\gamma\gamma}<110\,$GeV.  

The interaction of two photons can be classified
to be either a direct process where two bare photons interact, a single-resolved process
where a bare photon interacts with a parton of the other photon 
or a double-resolved process where partons of
both photons interact together. The last two processes are due to the possibility
of a photon to interact as a hadronic fluctuation. This classification is only uniquely 
defined in LO QCD, but not in NLO QCD.

In LO QCD, the experimental signature, neglecting multiple parton interactions,
consists of two hard parton jets with large transverse energy $E_{T}^{\rm jet}$ (direct events). 
In single- or double-resolved interactions, the
two hard parton jets are expected to be accompanied by one or two remnant jets.

The cone jet finding algorithm has been applied throughout the following studies
unless otherwise specified.

\begin{wrapfigure}{r}{8.2cm}
\hspace*{0.5cm}
\epsfig{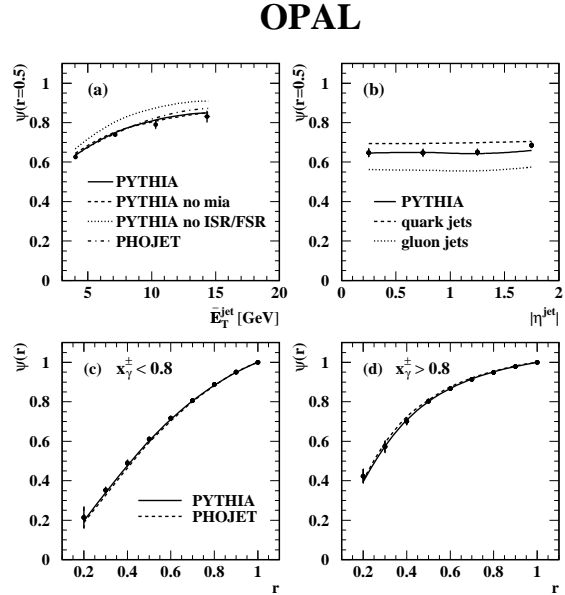}
\caption{\it Jet shape $\psi(r=0.5)$ as a function of $\bar{E}_{T}^{\rm jet}$ (a) and
$|\eta^{\rm jet}|$ (b). $\psi(r)$ as a function of $r$ for $x_{\gamma}^{\pm}<0.8$ (c)
and $x_{\gamma}^{\pm}>0.8$ (d).}
\label{fig14}
\end{wrapfigure}

The above three event classes can be separated using the fraction of the photon's momentum which 
participates in the hard interaction. It can be specified using the 
following relations \cite{ref13_lep2}:
\begin{equation}
x_{\gamma}^{\pm}=\frac{\sum_{{\rm jets=1,2}}(E\pm p_{z})}{\sum_{{\rm hadrons}}(E\pm p_{z})} 
\end{equation}
$p_{z}$ is the momentum component along the $z$ axis and $E$ is the energy 
of the respective energy depositions of the jets or hadrons.
These variables provide some separation of direct and resolved two-jet events \cite{ref27_opal}

The underlying processes for $\gamma\gamma$ scattering have been calculated in LO \cite{ref28_lo} and 
NLO perturbative QCD (pQCD) \cite{ref29_nlo}.
In the case of direct events, the only LO contributing matrix element is 
$\gamma\gamma \rightarrow \q \qb$ whereas for double-resolved events, quark-gluon, gluon-quark 
and gluon-gluon type matrix elements have to be taken into account. The observed jet events
are thus related to the underlying dynamics which allows to examine the structure the photon
and the dynamics of the $\gamma\gamma$ processes. 
The measured inclusive two-jet cross-section as well as the angular distribution
of the parton scattering angle in the two-jet centre-of-mass frame have been
compared to NLO pQCD calculations. 
These two items together with a study of jet shapes in two-jet events 
and an investigation of the influence of an underlying event in two-jet events will be
presented in the next section. The production of charged hadrons and
$\chi_{c2}$ mesons will be discussed in section 3.3. 
The measurement of the total hadronic
$\gamma\gamma$ cross-section will be focused on in detail in section 3.4.

\begin{wrapfigure}{r}{8.2cm}
\hspace*{0.5cm}
\epsfig{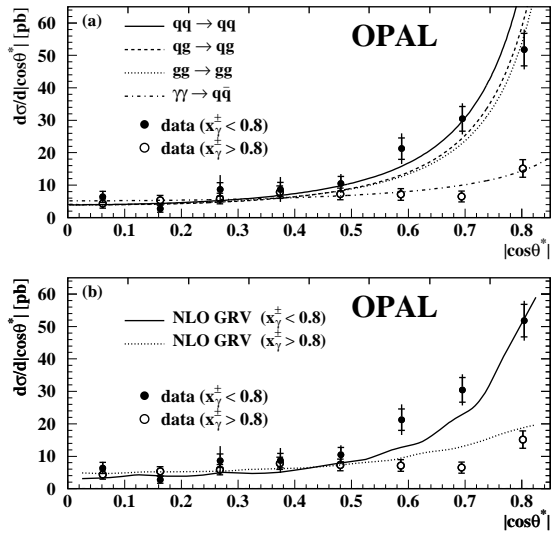}
\caption{\it Angular distribution $d\sigma/d|\cos\theta^{*}|$ as function
of $|\cos\theta^{*}|$ for $x_{\gamma}^{\pm}>0.8$ and $x_{\gamma}^{\pm}<0.8$
in comparison to NLO pQCD calculations.}
\label{fig15}
\end{wrapfigure}

\subsection{Jet production}

\subsubsection{Jet shapes}

The internal structure of jets produced in two-jet events has been studied based 
on the following jet shape definition:
\begin{equation}
\psi(r)=\frac{1}{N_{{\rm jet}}}\sum_{{\rm jet}}\frac{E_{T}(r)}{E_{T}(r=R)}
\end{equation}
$\psi(r)$ denotes the fraction of the jet's energy that lies inside
an inner cone of radius $r$, thus $\psi(r=R)=1$. 
$E_{T}(r)$ is the transverse energy within the inner
cone of radius $r$ and $N_{{\rm jet}}$ refers to the total number of jets in the sample.

Figure \ref{fig14} shows the fraction of the transverse energy of the jets
inside a cone of radius $r=0.5$ around the jet axis (a) as a function of 
$\bar{E}_{T}=(1/2)(E_{T}^{\rm jet1}+E_{T}^{\rm jet2})$ and (b) as
a function of $|\eta^{\rm jet}|=|(1/2)(\eta^{\rm jet1}+\eta^{\rm jet2})|$ 
with $E_{T}^{{\rm jet}\, i}$ and $\eta^{{\rm jet}\, i}$ defined in the lab frame.
The jet shapes have been corrected to the hadron level. 
The data points are
compared to PYTHIA with and without multiple interactions and to PYTHIA
with and without initial (ISR) and final state (FSR) QCD radiation along with
a prediction from PHOJET \cite{ref28_phojet}. 
Jets without initial and final
state QCD radiation are significantly narrower. Multiple
interactions as simulated within PYTHIA have only a minor effect on 
the jet shape. 
As expected, the jets become narrower with increasing $\bar{E}_{T}$.
No significant dependence of $\psi(r=0.5)$
on $|\eta^{\rm jet}|$ has been found. 

The gluon and quark
content in direct and double-resolved events is expected to be different
from the above discussion in section 3.1. 
Samples of large fractions of direct and double-resolved events have
been selected by requiring $x_{\gamma}^{\pm}$ to be larger 
or less than $0.8$, respectively. The jet shape $\psi(r)$ is shown in 
Figure \ref{fig14} as a function of $r$ for $x_{\gamma}^{\pm}<0.8$ (c)
and $x_{\gamma}^{\pm}>0.8$ (d). The observed jet shapes are found to be
broader for the double-resolved event sample ($x_{\gamma}^{\pm}<0.8$)
than those for direct events ($x_{\gamma}^{\pm}>0.8$). 
Gluon jets 
are known to be broader than quark jets which is consistent with the 
difference in the jet shapes of direct and double-resolved events
due to the expected difference in the gluon and quark composition. 

The PYTHIA and PHOJET jet shape prediction are found to be in good agreement 
with the data.

\subsubsection{Angular distributions in direct and resolved events}

The angular distribution for direct and double-resolved events in the two-jet
centre-of-mass frame is expected to be different due to the different 
gluon and quark composition and thus the different contributing matrix elements.
The angle $\cos\theta^{*}$ between the jet axis and the axis of the incoming partons
or direct photons
in the two-jet centre-of-mass frame can be estimated as follows:
\begin{equation}
\cos\theta^{*}=\tanh\left(\frac{\eta^{\rm jet1}-\eta^{\rm jet2}}{2}\right)
\end{equation}
In LO, the direct process $\gamma\gamma \rightarrow \q\qb$ leads to an 
angular dependence of the form $\propto 1/(1-\cos^{2}\theta^{*})$ whereas
for double-resolved events which involves gluon initiated matrix elements,
the angular dependence can be approximated as $\propto 1/(1-\cos^{2}\theta^{*})^{2}$.

\begin{wrapfigure}{r}{8.2cm}
\hspace*{0.75cm}
\vspace*{0.05cm}
\epsfig{figure=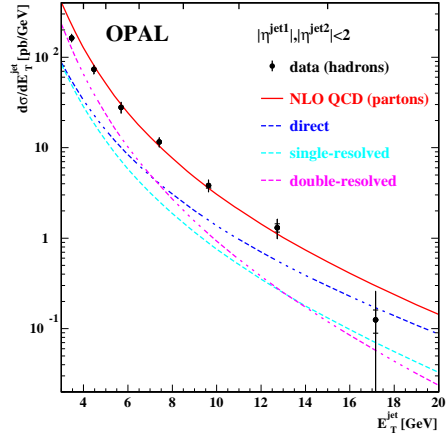,width=6.0cm,clip=}
\vspace*{-0.05cm}
\caption{\it The inclusive two-jet cross-section as a function of $E_{T}^{\rm jet}$ 
for $|\eta^{\rm jet}|<2$ in comparison to NLO pQCD calculations.}
\label{fig16}
\end{wrapfigure}

Figure \ref{fig15} shows the angular distribution $d\sigma/d|\cos\theta^{*}|$ as a function
of $|\cos\theta^{*}|$ for samples with a large fraction of direct ($x_{\gamma}^{\pm}>0.8$) and 
double-resolved ($x_{\gamma}^{\pm}<~0.8$) events. 
The data are compared to NLO pQCD calculations for
the cross-section $d\sigma/d|\cos\theta^{*}|$ and the contributing single cross-sections. 
Events with $x_{\gamma}^{\pm}>~0.8$ show only a small rise with increasing 
$|\cos\theta^{*}|$ whereas
events with $x_{\gamma}^{\pm}<0.8$ show a much larger rise as a function of $|\cos\theta^{*}|$. This
behavior follows the QCD expectation taking into account the expected difference in the 
gluon and quark composition of direct and double-resolved events. The result of the NLO
pQCD calculations agrees well with the observed shape of the data.

\subsubsection{Inclusive two-jet cross-section and NLO calculations}

Figure \ref{fig16} shows the inclusive two-jet cross-section as a function of $E_{T}^{\rm jet}$
for $|\eta^{\rm jet}|<2$ at $\sqrt{s_{\epl\emi}}=161-172\,$GeV.
The measurements are compared to NLO pQCD calculations by Kleinwort and Kramer \cite{ref29_kramer}
which are using the GRV parton parameterizations of the photon. The predictions for the 
direct, single- and double-resolved contributions as well as their sum are shown separately. The data
points are in good agreement with NLO pQCD calculations except for the first bin. 
However, experimental as well
as theoretical uncertainties are large in this kinematic region. The direct component dominates
the cross-section at high $E_{T}^{\rm jet}$, whereas the resolved cross-section is the largest
component for $E_{T}^{\rm jet}<8\,$GeV.

\subsubsection{Influence of the underlying event}

The possibility for an underlying event in the $\gamma\gamma$ scattering process is 
not taken
into account in the NLO pQCD calculations. However, an underlying event leads to an 
increased
jet cross-section. 
Resolved photons such as in double-resolved events
involve multiple partons which are subject to multiple interactions.
The effect of the simulation of the underlying event is therefore important. The
MC models PYTHIA and PHOJET use the concept of multiple interactions in the simulation of the underlying
event. 
The contribution of multiple interactions has to be tuned using quantities which are not
directly correlated to the jets. Only then effects due to parton distributions and
due to the underlying event can be disentangled. The model dependence for the simulation of 
the underlying event through multiple interactions can be tuned using the transverse cutoff parameter $p_{t}^{\rm mi}$
for multiple interactions.

\begin{wrapfigure}{r}{8.2cm}
\hspace*{0.05cm}
\vspace*{0.25cm}
\epsfig{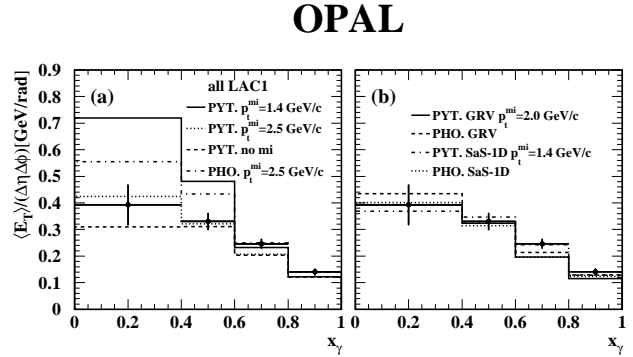}
\caption{\it Transverse energy flow outside the jets in the central rapidity region 
$|\eta^{*}|<1$ as a function of $x_{\gamma}$.}
\label{fig16}
\end{wrapfigure}

The investigation of an underlying event in the $\gamma\gamma$ scattering process
starts from the observation that the transverse energy flow outside the jets measured
as a function of $x_{\gamma}$ is correlated to the underlying event \cite{ref30_h1}. 
At small $x_{\gamma}$, the transverse energy flow outside the jets increases which can
be used to tune the number of multiple interactions.
The transverse energy flow corrected to the hadron level outside the jets, excluding
the region of $R<1.3$ in the energy sum, is shown in Figure \ref{fig16} as a function of $x_{\gamma}$
in comparison to different MC model predictions using different values for $p_{t}^{\rm mi}$.
The impact of multiple interactions has been found to be large for PYTHIA and PHOJET using 
the LAC1 parameterization. 
Their impact is small when using either GRV or SaS-1D. The 
optimal transverse cutoff parameter $p_{t}^{\rm mi}$ shows a strong dependence on the underlying
parton distributions in case of PYTHIA. Using GRV, $p_{t}^{\rm mi}$ has been set to $2.0\,$GeV,
SaS-~1D requires a value for $p_{t}^{\rm mi}$ of $1.4\,$GeV whereas for LAC1, $p_{t}^{\rm mi}$ has been
set to $2.5\,$GeV. In case of PHOJET the default cutoff parameter 
$p_{t}^{\rm mi}$ of $2.5\,$GeV 
provides for GRV as well as for SaS-1D a reasonable description of the transverse energy flow. 

\begin{wrapfigure}{r}{8.2cm}
\hspace*{0.05cm}
\epsfig{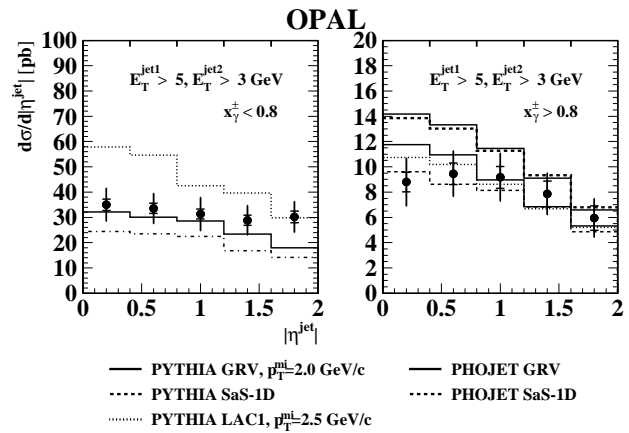}
\caption{\it Inclusive two-jet cross-section as a function of $|\eta^{\rm jet}|$ for
(left) $x_{\gamma}^{\pm}<0.8$ and (right) $x_{\gamma}^{\pm}>0.8$.}
\label{fig17}
\end{wrapfigure}

\subsubsection{Inclusive two-jet cross-section as a function of \boldmath{$|\eta^{\rm jet}|$}}

Having tuned the simulation of the underlying event through multiple interactions using the 
transverse cutoff parameter $p_{t}^{\rm mi}$, the sensitivity of the jet cross-section on the
underlying parton distribution can be investigated. The inclusive two-jet cross-section is 
shown in Figure~\ref{fig17} as a function of $|\eta^{\rm jet}|$ for $E_{T}^{\rm jet1}>5\,$GeV
and $E_{T}^{\rm jet2}>3\,$GeV with (left) large samples of double-resolved events ($x_{\gamma}^{\pm}<0.8$)
and (right) large samples of direct events ($x_{\gamma}^{\pm}>0.8$). 
A large dependence on the underlying
parton distribution is found for double-resolved events compared to direct events. The impact
of the gluon distribution is expected to be larger for double-resolved events compared to direct events. 
LAC1 compared to GRV and SaS-1D, which increases the cross-section for gluon-initiated 
processes, leads to a strong overestimation of the two-jet cross-section. 

\begin{wrapfigure}{r}{8.2cm}
\hspace*{0.75cm}
\epsfig{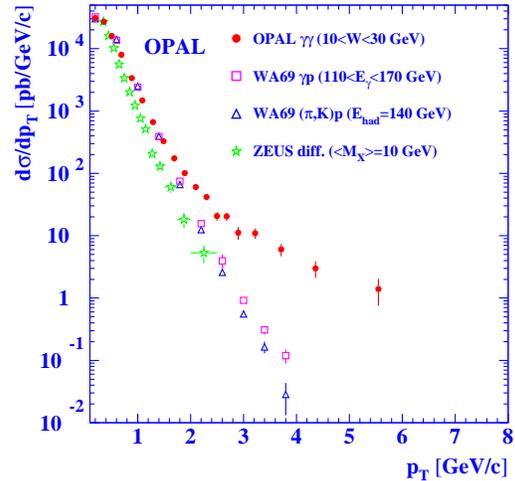}
\caption{\it $p_{T}$ distribution of $\gamma\gamma$, $\gamma \p$, $\pi \p$ and $\K \p$ scattering.} 
\label{fig18}
\end{wrapfigure}

\subsection{Hadron production}

\subsubsection{Production of charged hadrons}

The contribution of direct photon processes in photon induced interactions 
is expected to lead to a harder transverse momentum distribution of charged hadrons
for $\gamma\gamma$ scattering
than for $\gamma \p$ or hadron-$\p$ scattering. The $p_{T}$ distribution for $\gamma\gamma$ scattering
as obtained by OPAL \cite{ref31_opal} with $10<W<30\,$GeV is shown in Figure \ref{fig18} together with data
on $\gamma \p$, $\pi \p$ and $\K \p$ data from WA69 with a hadronic invariant mass of $16\,$GeV. 
The WA69 data are normalized to the $\gamma\gamma$
data at $p_{T}\approx 200\,$MeV. ZEUS data on charged particle production in
$\gamma \p$ scattering with a diffractively dissociated photon, having an average invariant 
mass of the `$\gamma$-Pomeron' system of $10\,$GeV, 
are shown as well. 

It can be clearly inferred from Figure \ref{fig18} that $\gamma\gamma$ interactions show a significantly
harder $p_{T}$ distribution than $\gamma \p$ and hadron-$\p$ interactions. The measured 
$p_{T}$ distribution has been compared to NLO pQCD calculations which have been found to be in good 
agreement with the data. The direct interactions yield the largest contribution at high $p_{T}$
compared to single- and double-resolved interactions.

\begin{wrapfigure}{r}{8.2cm}
\hspace*{0.75cm}
\epsfig{figure=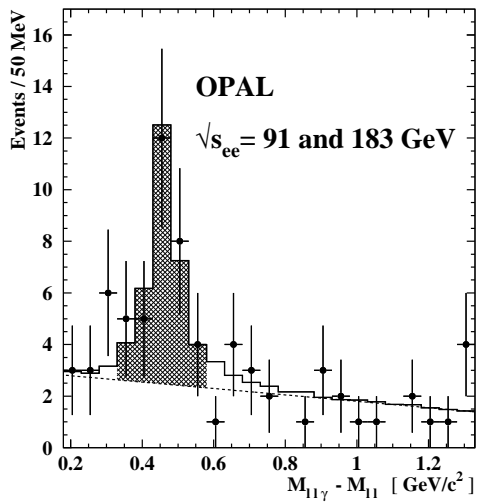,width=6.5cm,clip=}
\caption{\it Mass difference between the $\lep\lep\gamma$ and $\lep\lep$ invariant masses, 
$M_{\lep\lep\gamma}-M_{\lep\lep}$.}
\label{fig19}
\end{wrapfigure}

\subsubsection{Production of \boldmath{$\chi_{c2}$} mesons} 

OPAL has carried out an analysis reconstructing $\chi_{c2}$ mesons in the decay channel 
$\chi_{c2}\rightarrow J/\psi\;\gamma \rightarrow \lpl\lmi\gamma \;\;\;(\lep=\e,\mu)$ 
which resulted in a measurement of the two-photon width $\Gamma(\chi_{c2}\rightarrow \gamma\gamma)$
\cite{ref32_opal}.
The data sample consists of all data taken at $\epl\emi$ centre-of-mass energies of $91\,$GeV 
and $183\,$GeV corresponding to integrated luminosities of $167\,$pb$^{-1}$ and $55\,$pb$^{-1}$, respectively.
Figure \ref{fig19} shows the mass difference between the $\lep\lep\gamma$ and $\lep\lep$ invariant masses, 
$M_{\lep\lep\gamma}-M_{\lep\lep}$. 
$34$ events have been selected in the $\chi_{c2}$ signal region of $330<M_{\lep\lep\gamma}-M_{\lep\lep}<550\,$MeV 
including a background of $12.4\pm 3.3$ events.
The contribution from $\chi_{c0}$ and $\chi_{c1}$ is estimated not to exceed a few percent.
The two-photon width has been determined to be 
$\Gamma(\chi_{c2}\rightarrow \gamma\gamma)=1.76 \pm 0.47 \pm 0.37 \pm 0.15\,$keV. The first error
denotes the statistical and the second the systematic error. The third error reflects the uncertainty of the 
branching
ratios in the decay $\chi_{c2}\rightarrow J/\psi\;\gamma \rightarrow \lpl\lmi\gamma$. This result 
agrees with results from CLEO, TPC/2$\gamma$ and R704, but is about two standard deviations larger 
than the E760 result and the current world average \cite{ref33_barnett}. 
Furthermore the determined two-photon width is found to be two standard deviations larger than the 
prediction by Schuler \cite{ref34_schuler}. 

\begin{figure}[ht]
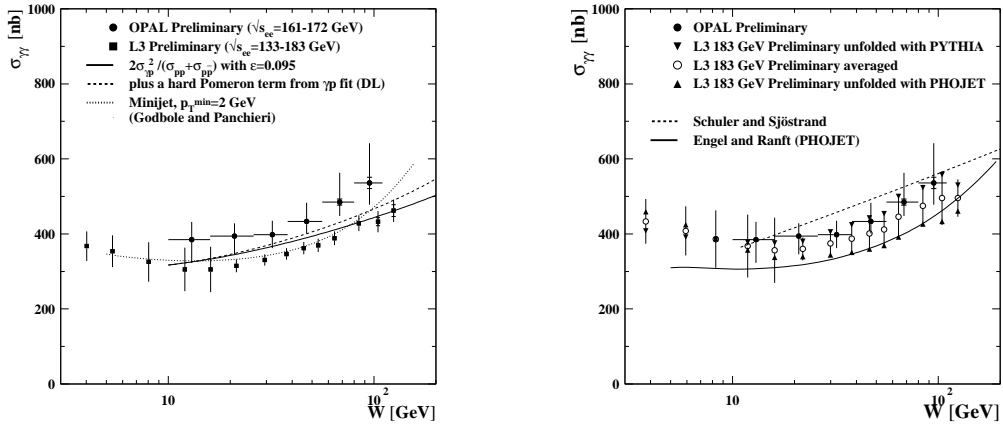

\setlength{\unitlength}{1.0cm}
\begin{picture} (15.0,6.0
)
\put (1.25,0.0){\mbox{\psfig{figure=tot_bw2.eps,width=6.0cm,clip=}}}
\put (8.75,0.0){\mbox{\psfig{figure=tot_bw3.eps,width=6.0cm,clip=}}}
\end{picture}
\caption{\it Total cross-section for $\gamma\gamma\rightarrow {\rm hadrons}$, 
$\sigma_{\gamma\gamma}$, as 
a function of $W=\sqrt{s_{\gamma\gamma}}$ from OPAL in comparison to L3 results and
theoretical predictions (left). Comparison of preliminary OPAL results on $\sigma_{\gamma\gamma}$
with L3 results obtained with PYTHIA and PHOJET (right) \cite{ref3_soldner}.}
\label{fig20}
\end{figure}

\subsection{Total cross-sections}

The total cross-section for $\gamma\gamma \rightarrow \; {\rm hadrons}$, $\sigma_{\gamma\gamma}$,
at large centre-of-mass energies $W=\sqrt{s_{\gamma\gamma}}$ 
is expected to be dominated by interactions of hadronic fluctuations of
the colliding photons. A measurement of $\sigma_{\gamma\gamma}$ should therefore 
improve our understanding of the hadronic nature of the photon and in particular
clarify the question whether the $W$ dependence of $\sigma_{\gamma\gamma}$ exhibits the same behavior
as total hadronic cross-sections. This could lead to an answer of the question
if there is an universal behavior of total cross-sections in $\gamma\gamma$, $\gamma \p$ and 
$\p\p$/$\p\bar{\p}$ 
scattering. 

Pre-Lep data are restricted to rather 
low energies in $W$, which is too low to observe
the expected rise of total hadronic cross-sections due to Pomeron exchange in
the language of Regge phenomenology. 
A measurement at LEP taking into account the large centre-of-mass energy compared
to previous experiments at PETRA allow to study the behavior of $\sigma_{\gamma\gamma}$
at higher energies.

The extraction of the total cross-section, $\sigma_{\gamma\gamma}$,
relies heavily on a particular Monte Carlo (MC) program to be used for the unfolding. 
It is therefore
expected that systematic uncertainties on the extracted total cross-sections are
dominated by the simulation of the hadronic final state. 

Figure \ref{fig20} (left) shows the preliminary OPAL results \cite{ref35_opal} 
on the total cross-section, $\sigma_{\gamma\gamma}$, as
a function of $W=\sqrt{s_{\gamma\gamma}}$ in the range of $10 \leq W \leq 110\,$GeV 
in comparison to L3 results \cite{ref36_l3} for $5 \leq W \leq 145\,$GeV. 
Various curves from theoretical predictions are overlaid. 
The preliminary OPAL cross-section results are quoted 
by taking the average of the cross-section results obtained with the MC models
PHOJET and PYTHIA. The difference between theses results has been
included in the systematic error evaluation.
However, L3 has used PHOJET to determine the central values. No
model dependence is reflected in their quoted systematic errors. The unfolded L3
results using PHOJET and PYTHIA together with their average values are shown in Figure
\ref{fig20} (right). Comparing theses cross-section values with the preliminary results obtained
by OPAL and thus providing a consistent bases for a detailed comparison, no significant
discrepancy between both results is found. 

Relating the total cross-section, $\sigma_{\gamma\gamma}$, to the total cross-sections
for $\gamma \p$, $\p\p$ and $\bar{\p}\p$ scattering, 
results in the prediction shown as the solid line
in Figure \ref{fig20} (left). 
It describes the general trend of $\sigma_{\gamma\gamma}$ as a function of $W$.
It has been argued by many authors to expect a faster rise of the total cross-section
for $\gamma\gamma$ scattering compared to $\gamma \p$ scattering and hadron-hadron scattering
due to the possibility of two photons interacting directly. In a Regge-type fit to the data,
L3 determined a Pomeron intercept of 
$\pom=1.158 \pm 0.006 \, ({\rm stat}) \pm 0.028 \, ({\rm sys})$
which is 
larger than the Pomeron intercept of 
$\pom=1.100 \pm 0.002 \, ({\rm stat}) \pm 0.012 \, ({\rm sys})$
obtained from a recent ZEUS analysis of total
$\gamma \p$ cross-section results based on an extrapolation of total
$\gamma^{*} \p$ cross-section data \cite{ref37_zeus}. The soft Pomeron intercept by 
Donnachie and Landshoff 
\cite{ref_dl} of $\pom=1.0808$ is smaller than those obtained from total-cross section
results involving at least one photon in the scattering process. 

\section{Summary and Outlook}

OPAL has performed a wide range of measurements of $\e\gamma$ and $\gamma\gamma$ 
scattering in $\epl\emi$ collisions at LEP. 

The hadronic photon structure function
$F_{2}^{\gamma}$ has been measured for 
$0.0025<x \lsim 1$ and 
$1.1<Q^{2}<400\,{\rm GeV}^{2}$. The current systematic
uncertainties on $F_{2}^{\gamma}$ are dominated by the model dependence
in the simulation of the hadronic final state. Only considerable
improvements of general purpose Monte Carlo (MC) models to 
better describe the hadronic final state will allow 
to study the low $x$ behavior of $F_{2}^{\gamma}$ as well as 
the $Q^{2}$ evolution with higher precision. Data on the leptonic structure function
$F_{2, {\rm QED}}^{\gamma}$ are precise enough to test QED and to be sensitive
to more subtle effects such as the $P^{2}$ dependence. 

It has been shown that the measured inclusive two-jet cross-sections in $\gamma\gamma$ scattering 
can be well understood within the framework of NLO perturbative QCD (pQCD). 
Two-jet cross-section 
measurements provide
a useful means to constrain the gluon density in the photon. The 
total cross-section results of the process $\gamma\gamma \rightarrow {\rm hadrons}$ from OPAL
and L3 were shown to be consistent. It has been found that the total $\gamma\gamma$
cross-section rises faster with $W$ than in $\gamma \p$ scattering and
even faster than in $\p\p$ or $\p\bar{\p}$ scattering.

The main concern for future measurements of $F_{2}^{\gamma}$ and
$\sigma_{\gamma\gamma}$ is to reduce the large systematic uncertainties 
due to the simulation of the hadronic final state. Improvements of general purpose
MC models are eagerly awaited \cite{ref18_leif}.
With the larger centre-of-mass energy $\sqrt{s_{\epl\emi}}$ at LEP2 and the higher
statistics it will be possible to extend the kinematic range towards larger values in 
$Q^{2}$ and $W^{2}$.
Finally, the study of double-tagged events to investigate the virtual structure of the
photon will be one of the main analysis topics for the future. 

An exciting time is ahead of us until 2000 investigating the structure of the
photon and its interactions at LEP2\footnote{At the time of writing this report, the
OPAL experiment at LEP has recorded more than $175\,$pb$^{-1}$ of data during the '98 run 
at a centre-of-mass energy of $189\,$GeV.}.

\pagebreak

\begin{flushleft}
{\Large\bf Acknowledgement}
\end{flushleft}

I want to thank Richard Nisius, Gerhard Schuler and Stefan S\"{o}ldner-Rembold 
for carefully reading the manuscript. I am very grateful to Goran Jarlskog
for giving me the opportunity to go to this excellent workshop.

\end{document}